\documentclass[aps,prx,preprint,nofootinbib, superscriptaddress]{revtex4-2}

\usepackage{amsmath,amssymb}
\usepackage{graphicx}
\usepackage{color}
\usepackage{rotating}
\usepackage{bm}
\usepackage{setspace}
\usepackage{hyperref}
\usepackage{multirow}

\hypersetup{
colorlinks=true,
urlcolor= blue,
citecolor=blue,
linkcolor= blue,
bookmarks=true,
bookmarksopen=false,
}

\begin{document}


\title{Substrate tuning of the structural and electronic transition \\ in thin flakes of the excitonic insulator candidate Ta$_2$NiSe$_5$}


\author{Yuan-Shan Zhang$^a$}
\email{Yuanshan.Zhang@fkf.mpg.de}
\affiliation{Max Planck Institute for Solid State Research, 70569 Stuttgart, Germany}
\author{Zichen Yang$^a$}
\affiliation{Max Planck Institute for Solid State Research, 70569 Stuttgart, Germany}
\author{Chuanlian Xiao}
\affiliation{Max Planck Institute for Solid State Research, 70569 Stuttgart, Germany}
\author{Masahiko Isobe}
\affiliation{Max Planck Institute for Solid State Research, 70569 Stuttgart, Germany}
\author{Matteo Minola}
\affiliation{Max Planck Institute for Solid State Research, 70569 Stuttgart, Germany}
\author{Hidenori Takagi}
\affiliation{Max Planck Institute for Solid State Research, 70569 Stuttgart, Germany}
\affiliation{Institute for Functional Matter and Quantum Technologies, University of Stuttgart, 70569 Stuttgart, Germany}
\affiliation{Department of Physics, University of Tokyo, 113-0033 Tokyo, Japan}
\author{Dennis Huang}
\email{D.Huang@fkf.mpg.de}
\affiliation{Max Planck Institute for Solid State Research, 70569 Stuttgart, Germany}

\date{\today}

\begin{abstract}
Ta$_2$NiSe$_5$ continues to draw interest for its $T_\textrm{c}$ = 326 K phase transition, whose dual electronic and structural nature reflects a complex interplay of electron-hole (excitonic) and electron-lattice interactions. The majority of studies that have attempted to decipher the relative importance of these interactions, particularly through charge tuning, have been focused on bulk samples. Here, we utilize an all-dry exfoliation and transfer protocol to isolate ultrathin flakes of Ta$_2$NiSe$_5$ on insulating Al$_2$O$_3$ and conducting Au. Using polarized Raman spectroscopy, we uncover the following substrate dependence: Four layers of Ta$_2$NiSe$_5$ on Al$_2$O$_3$ show a sharp structural and electronic transition that is lowered by roughly 40 K from the bulk $T_\textrm{c}$. Meanwhile, four layers of Ta$_2$NiSe$_5$ on Au undergo a structural and electronic transition that is much more gradual with respect to temperature and finishes roughly 150 K below the bulk $T_\textrm{c}$. The pronounced broadening points to an atomic-scale interface effect, wherein electrostatic screening and charge transfer from Au produces a $T_\textrm{c}$ gradient perpendicular to the layers of the flake. We discuss the role of excitonic physics and suggest the possibility for interface engineering to pattern nanoscale junctions in Ta$_2$NiSe$_5$.  
\end{abstract}


\maketitle
\def\thefootnote{$a$}\footnotetext{equal contribution}

\newpage\section*{Introduction}

The excitonic insulator (EI) is a macroscopic condensate of electron-hole pairs, i.e., excitons, predicted to arise near the semiconductor-semimetal boundary, when the exciton binding energy exceeds the narrow band gap or overlap \cite{Mott_1961, Jerome_1967, Halperin_Rice_RMP_1968}. Despite the apparent analogy to superconductivity, which is a macroscopic condensate of electron-electron pairs, the EI has proven to be much more elusive as the ground state of any bulk crystal. One inherent difficulty with an EI is that the electrons and holes originate from different bands, which often reside in distinct regions of the Brillouin zone or comprise distinct orbital characters. An equilibrium interaction of electrons and holes with incompatible Bloch wave functions necessitates a structural transition that lowers the translational and/or point group symmetries of the lattice. The coupling of electrons and holes not only to each other, but also to the lattice, considerably enriches the concept of an EI and makes it an attractive focus of study in strongly correlated physics.

Ta$_2$NiSe$_5$ exemplifies this EI challenge. It is a quasi-one-dimensional chalcogenide comprising chains of TaSe$_6$ octahedra and NiSe$_4$ tetrahedra running along the $a$ axis, with successive layers weakly stacked along the $b$ axis (Fig.~\ref{Fig1}a) \cite{Sunshine_1985, DiSalvo_1986}. The valence band maximum and conduction band minimum lie directly on top of each other at the $\Gamma$ point, and are composed of Ni 3$d$/Se 4$p$ and Ta 5$d$ orbitals, respectively \cite{Kaneko_PRB_2013}. Far above room temperature, the resistivity of Ta$_2$NiSe$_5$ is nearly independent of temperature, suggestive of poor metallic behavior in its normal state. Below $T_{\textrm{c}}$ = 326 K, the resistivity shows insulating behavior with the corresponding opening of an optical gap with value between 0.16 eV (full gap) and 0.3 eV (isosbestic point), close to the exciton binding energy observed in Ta$_2$NiS$_5$ \cite{DiSalvo_1986, Lu_NatComm_2017, Larkin_2017}. Early reports pointed to an apparent flattening of the valence band below $T_{\textrm{c}}$ \cite{Wakisaka_2009}, as well as a dome-like dependence of $T_{\textrm{c}}$ on the normal-state band gap/overlap \cite{Lu_NatComm_2017}, as hallmarks of a canonical EI with a many-body charge gap. 

At the same time, Ta$_2$NiSe$_5$ undergoes a shearing distortion of neighboring Ta-ion chains below $T_{\textrm{c}}$, which preserves the volume of the unit cell, but lowers its symmetry from orthorhombic to monoclinic \cite{DiSalvo_1986, Nakano_PRB_2018}. The monoclinic distortion lifts a mirror plane and allows Ni 3$d$ and Ta 5$d$ orbitals with distinct symmetries from the valence and conduction bands to hybridize \cite{Watson_PRR_2020, Mazza_2020, Subedi_2020}. The system can thus gain electronic energy by opening up a hybridization gap in cooperation with a lattice distortion, which is analogous to a $Q$ = 0 charge density wave. This hybridization gap enabled by monoclinic symmetry not only mixes single-particle character into the charge gap of Ta$_2$NiSe$_5$, but also has been argued to imply a crucial role of electron-lattice coupling in the 326 K transition.

A basic experimental strategy for disentangling the contributions of electron-hole (excitonic) and electron-lattice interactions in Ta$_2$NiSe$_5$ involves applying various kinds of perturbation, including external pressure and strain \cite{Lu_NatComm_2017, Matsubayashi_2021, Okamura_2023, Shi_PRL_2025, Rosenberg_PRL_2026}, S substitution of Se \cite{Lu_NatComm_2017, Ye_2021, Volkov_PRB_2021, Saha_2022, Chen_NatComm_2023, Li_2024}, charge doping \cite{Hirose_2023, Song_2023, Fukutani_PRL_2019, Chen_PRB_2020, Chen_PRR_2023, Lee_ACS_Nano_2024, He_2021}, and optical pumping \cite{Mor_2017, Werdehausen_2018, Bretscher_SciAdv_2021, Bretscher_NatComm_2021, Golez_2022, Katsumi_2023, Baldini_PNAS_2023, Haque_2024}, and monitoring the resulting changes in the structural and electronic transition. The effect of doping is anticipated to be especially revealing, as additional charges directly tune the excitonic interactions by rapidly screening out the Coulomb attraction that binds electron-hole pairs together \cite{Jerome_1967, Zittartz_1967}. Common methods for charge tuning include doping bulk crystals of Ta$_2$NiSe$_5$ through chemical substitution \cite{Hirose_2023, Song_2023}, or doping just the surface of a Ta$_2$NiSe$_5$ crystal, either with K adatoms \cite{Fukutani_PRL_2019, Chen_PRB_2020, Chen_PRR_2023, Lee_ACS_Nano_2024}, or with the tip of a scanning tunneling microscope \cite{He_2021}. An alternative route to charge tuning Ta$_2$NiSe$_5$ is inspired by the field of two-dimensional (2D) van der Waals (vdW) semiconductors, where much success with modifying exciton binding energies in $M$$X_2$ ($M$ = Mo, W; $X$ = S, Se) has been obtained by thinning the compounds down to the atomic limit and interfacing them with various substrates \cite{Chernikov_PRL_2014, Lin_2014, Buscema_2014, He_PRL_2014, Raja_2017, Pan_2022}. The reduction of layers generally decreases interlayer electrostatic screening, but at the same time, causes the system to be more susceptible to environmental electrostatic screening. Conducting substrates can strongly suppress exciton binding energies in $M$$X_2$ via screening and/or charge transfer, without having to introduce more disorder or strongly perturb the lattice. In principle, the weak vdW stacking of Ta$_2$NiSe$_5$ along the $b$ axis allows thin flakes to be prepared \cite{Kim_ACSNano_2016, Li_2016, Qiao_2021, Zhang_2021, Liu_2024, Wei_NatComm_2025}, but the possible influence of conducting substrates remains to be investigated.      

Here, we apply evaporation-assisted exfoliation and introduce a transfer technique based on cold welding to fabricate Ta$_2$NiSe$_5$ thin flakes on insulating Al$_2$O$_3$ and conducting Au. We employ polarizer-resolved Raman spectroscopy to track the structural and electronic $T_{\textrm{c}}$ of the flakes with varying thicknesses on the two substrates. On Al$_2$O$_3$, four layers of Ta$_2$NiSe$_5$ maintain a sharp transition, whose $T_{\textrm{c}}$ is moderately reduced by $\sim$40 K from the bulk value, implying minimal influence from the substrate. On Au, four layers of Ta$_2$NiSe$_5$ exhibit a structural and electronic transition that is significantly broadened and finishes $\sim$150 K below the bulk value. This behavior unveils an interface effect originating from the Au substrate, due to a combination of screening and charge transfer: The Ta$_2$NiSe$_5$ layers closest to Au have a local $T_{\textrm{c}}$ reduced by $\sim$150 K, whereas the surface layers have a local $T_{\textrm{c}}$ closer to room temperature. Extraction of parameters from the cross-polarized Raman susceptibility related to electron-phonon coupling and critical excitonic fluctuations suggests that the latter is more directly impacted by the Au substrate.

\section*{Results}

\subsection*{All-dry exfoliation and transfer of thin Ta$_2$NiSe$_5$ flakes}

\begin{figure}
\includegraphics[width=\textwidth]{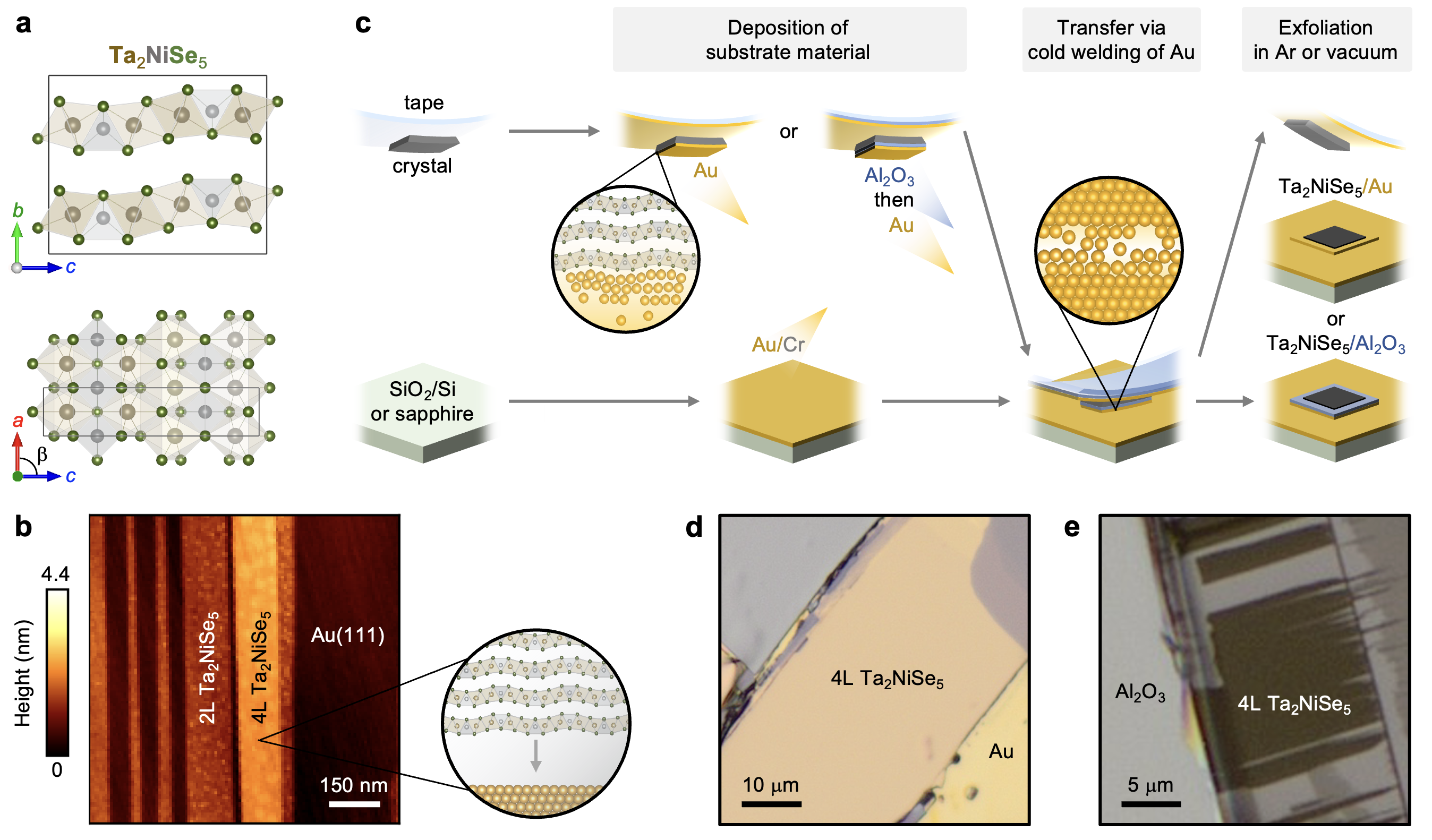}
\caption{\textbf{Preparation of ultrathin Ta$_2$NiSe$_5$ flakes.} \textbf{a} Crystal structure of Ta$_2$NiSe$_5$. The unit cell comprises two Ta$_2$NiSe$_5$ layers (L). In the orthorhombic phase, $\beta = 90^{\circ}$, whereas in the monoclinic phase at 30 K, $\beta \approx 90.7^{\circ}$ \cite{Nakano_PRB_2018}. \textbf{b} Topography of Ta$_2$NiSe$_5$ ribbons exfoliated onto an epitaxial Au(111) film on mica, acquired with atomic force microscopy. The inset cartoon illustrates the challenge of exfoliating a rippled compound onto an atomically flat surface. \textbf{c} Exfoliation and transfer technique assisted by thermal evaporation of the substrate material and cold welding of Au. \textbf{d}, \textbf{e} Optical micrographs of Ta$_2$NiSe$_5$ thin flakes on amorphous Au and Al$_2$O$_3$, respectively.}
\label{Fig1}
\end{figure}

Using the conventional Scotch tape method, we found it difficult to exfoliate Ta$_2$NiSe$_5$ flakes with suitable morphology below 10 nm thicknesses \cite{Bretscher_SciAdv_2021, Qiao_2021, Zhang_2021, Liu_2024}. The rare flakes that were as thin as 1 or 2 unit cells, i.e., 2 or 4 layers (L), tended to fragment into long ribbons with submicron widths (Fig.~\ref{Fig1}b) \cite{Li_NL_2024}. We surmise that the rippled nature of Ta$_2$NiSe$_5$ layers, which arises from the alternating arrangement of NiSe$_4$ tetrahedra and TaSe$_6$ octahedra, reduces the contact area and overall adhesion of Ta$_2$NiSe$_5$ with a flat substrate, such as Au(111)/mica (Fig.~\ref{Fig1}b inset). To improve the adhesion, we instead deposited either Au \cite{Huang_NatComm_2020} or Al$_2$O$_3$ \cite{Deng_Nature_2018} as the substrate directly onto a freshly cleaved crystal of Ta$_2$NiSe$_5$, which was fixed on a piece of tape. During the deposition process, the Au or Al$_2$O$_3$ adatoms likely conform better to the surface ripples and bind more strongly to Ta$_2$NiSe$_5$. To flip and transfer the Au- or Al$_2$O$_3$-coated Ta$_2$NiSe$_5$ onto a wafer, we developed a procedure based on the cold-welding ability of Au \cite{Lu_NatNano_2010}, as shown in Fig.~\ref{Fig1}c. We deposited Au onto a wafer and onto the Ta$_2$NiSe$_5$ (which is either bare or covered with Al$_2$O$_3$ from the previous step), then gently pressed the two Au-coated surfaces together immediately after deposition. The contact causes the Au-Au interface to bind strongly, thereby shifting the weakest interface of the entire structure to the Ta$_2$NiSe$_5$ layers. We peeled off the tape, leaving behind a high density of thin Ta$_2$NiSe$_5$/Au or Ta$_2$NiSe$_5$/Al$_2$O$_3$ flakes on the wafer \textcolor{blue}{(see Methods and Supplementary Note 1 with Supplementary Figure 1 for further fabrication details)}. 

We could regularly isolate few-layers-thick Ta$_2$NiSe$_5$ flakes with lateral dimensions of 10--100 $\mu$m using this method of evaporation and cold-welding (Fig.~\ref{Fig1}d, e). An advantage of Au cold-welding is that it is a dry process, which can be applied universally after the deposition of any substrate onto Ta$_2$NiSe$_5$. In contrast, we found that the use of solvents \cite{Huang_NatComm_2020} to transfer Ta$_2$NiSe$_5$ flakes between different substrates degraded the quality of the flakes. Our entire procedure can be performed in an inert Ar environment; the last step of removing the tape can also be carried out in ultra-high vacuum.

\subsection*{Polarized Raman spectroscopy}

\begin{figure}
\includegraphics[width=\textwidth]{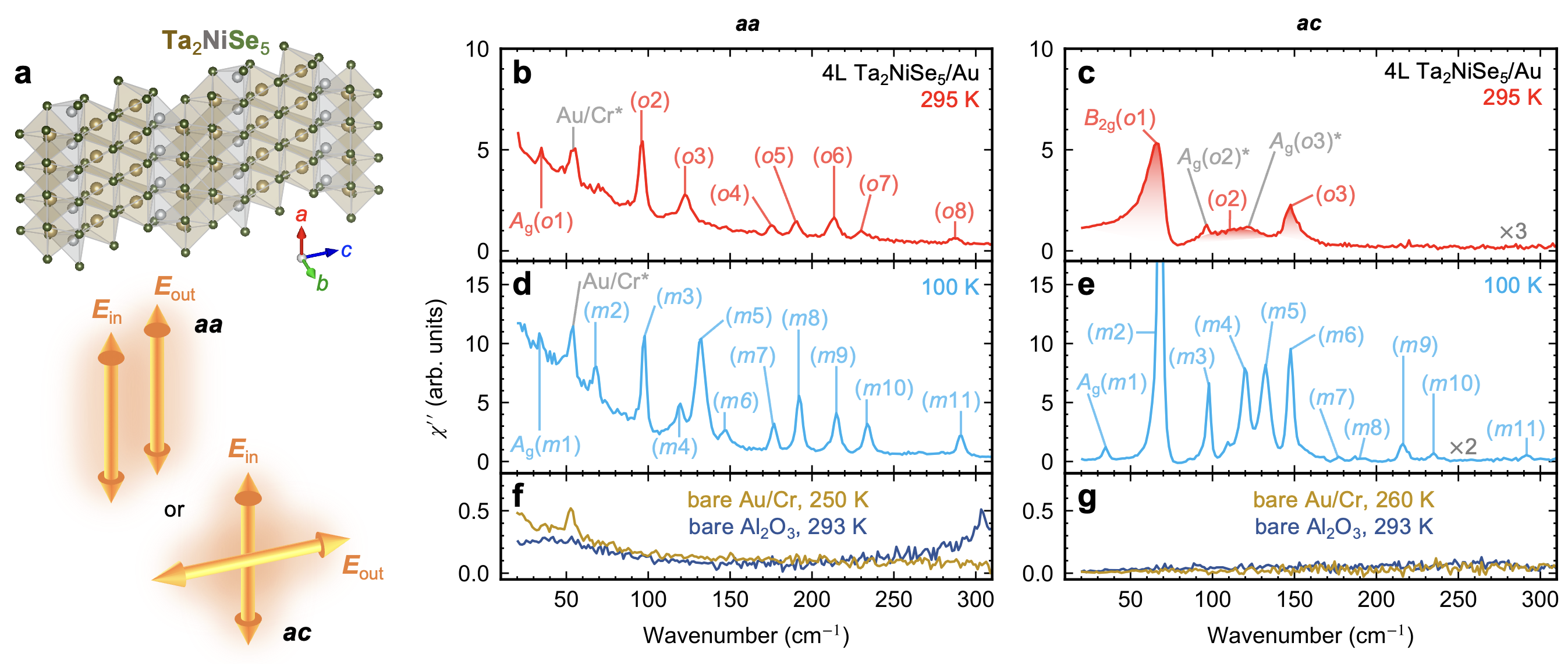}
\caption{\textbf{Raman-active phonon modes of Ta$_2$NiSe$_5$ thin flakes.} \textbf{a} Parallel ($aa$) and cross ($ac$) configurations of polarizer and analyzer employed for Raman spectroscopy. \textbf{b}, \textbf{c} The $aa$- and $ac$-Raman susceptibilities ($\chi''$) of a 4L Ta$_2$NiSe$_5$ flake on Au, acquired at 295 K. The phonon modes $A_{\textrm{g}}$($o$1)--$A_{\textrm{g}}$($o$8) and $B_{2\textrm{g}}$($o$1)--$B_{2\textrm{g}}$($o$3) of the orthorhombic phase are labeled. The red shading indicates the widened range of the $B_{2\textrm{g}}$ modes. The signals marked by asterisks originate either from the substrate (Au/Cr), or from a slight misalignment of the polarizer and analyzer, resulting in a leakage of modes $A_{\textrm{g}}$($o$2) and $A_{\textrm{g}}$($o$3) into the $ac$ channel. \textbf{d}, \textbf{e} The $aa$- and $ac$-Raman susceptibilities of the same flake, acquired at 100 K. The phonon modes $A_{\textrm{g}}$($m$1)--$A_{\textrm{g}}$($m$11) of the monoclinic phase are labeled. Note that the spectra in (\textbf{c}) and (\textbf{e}) have been scaled by the indicated factors for visualization. \textbf{f}, \textbf{g} The $aa$- and $ac$-Raman susceptibilities of bare Au/Cr and Al$_2$O$_3$.}
\label{Fig2}
\end{figure}

Figure \ref{Fig2}a illustrates the geometry of our Raman measurements: Incident photons were linearly polarized along the $a$ axis (chain direction) and propagated along the $b$ axis (interlayer direction); scattered photons were collected along the $b$ axis and were linearly analyzed either parallel or perpendicular to the $a$ axis, yielding two symmetry channels, parallel $aa$ and cross $ac$. Figure \ref{Fig2}b, c displays the $aa$- and $ac$-Raman susceptibilities ($\chi''$) of a 4L Ta$_2$NiSe$_5$ flake on Au at 295 K. We identified 11 Raman-active phonon modes, eight in the $aa$ channel and three in the $ac$ channel (red labels), which were previously observed in bulk Ta$_2$NiSe$_5$ \cite{Kim_NatComm_2021, Volkov_2021}. By rotating the sample relative to the polarizer and mapping out the azimuthal dependence of the mode intensities, we confirmed the eight modes in the $aa$ channel to have $A_{\textrm{g}}$ symmetry and the three modes in the $ac$ channel to have $B_{\textrm{2g}}$ symmetry, consistent with the selection rules \cite{Rousseau_1981} for the orthorhombic ($o$) phase of Ta$_2$NiSe$_5$ with $Cmcm$ space group \textcolor{blue}{(see Supplementary Note 2 with Supplementary Table 1 and Supplementary Figures 2 and 3 for further details)}. Modes $A_{\textrm{g}}$($o$1)--$A_{\textrm{g}}$($o$8) have symmetric and narrow line shapes (Fig.~\ref{Fig2}b), whereas modes $B_{2\textrm{g}}$($o$1) and $B_{2\textrm{g}}$($o$2) are noticeably widened over a range of $\sim$50 cm$^{–1}$, and modes $B_{2\textrm{g}}$($o$1) and $B_{2\textrm{g}}$($o$3) are also strongly asymmetric (Fig.~\ref{Fig2}c). 

Figure \ref{Fig2}d, e displays $\chi''_{aa}$ and $\chi''_{ac}$ of the same flake at 100 K. Here, all 11 Raman-active phonon modes are detectable in both the $aa$ and $ac$ channels, which is consistent with the change in selection rules for the monoclinic ($m$) phase of Ta$_2$NiSe$_5$ with $C2/c$ space group. We rename these modes $A_{\textrm{g}}$($m$1)--$A_{\textrm{g}}$($m$11) (blue labels), as $B_{2\textrm{g}}$ symmetry in the monoclinic unit cell is indistinguishable from $A_{\textrm{g}}$. The line shapes of these modes have also become narrower and more symmetric. We confirm that a 4L flake of Ta$_2$NiSe$_5$ on Au undergoes a structural transition from orthorhombic to monoclinic unit cells upon cooling, similar to bulk Ta$_2$NiSe$_5$.

Since the penetration depth of photons exceeds the thickness of a few Ta$_2$NiSe$_5$ layers, we performed control measurements with the beam spot positioned over the bare substrate. As seen in Fig.~\ref{Fig2}f, $\chi''_{aa}$ of amorphous Au (with Cr adhesion layer underneath) exhibits a narrow mode around 50 cm$^{-1}$, as well as a wide electronic background below $\sim$150 cm$^{-1}$. Both features are inherited by $\chi''_{aa}$ of 4L Ta$_2$NiSe$_5$/Au (Fig.~\ref{Fig2}b, d). In contrast, $\chi''_{ac}$ of Au/Cr is featureless (Fig.~\ref{Fig2}g), meaning that we can exclude substrate contributions to $\chi''_{ac}$ of 4L Ta$_2$NiSe$_5$/Au (Fig.~\ref{Fig2}c). The $ac$ configuration is similarly effective at suppressing Raman signals from bare Al$_2$O$_3$ (Fig.~\ref{Fig2}f, g). We therefore focus on $\chi''_{ac}$ in the subsequent analyses. 

\subsection*{Substrate and layer dependence of structural transition}

\begin{figure}
\includegraphics[width=\textwidth]{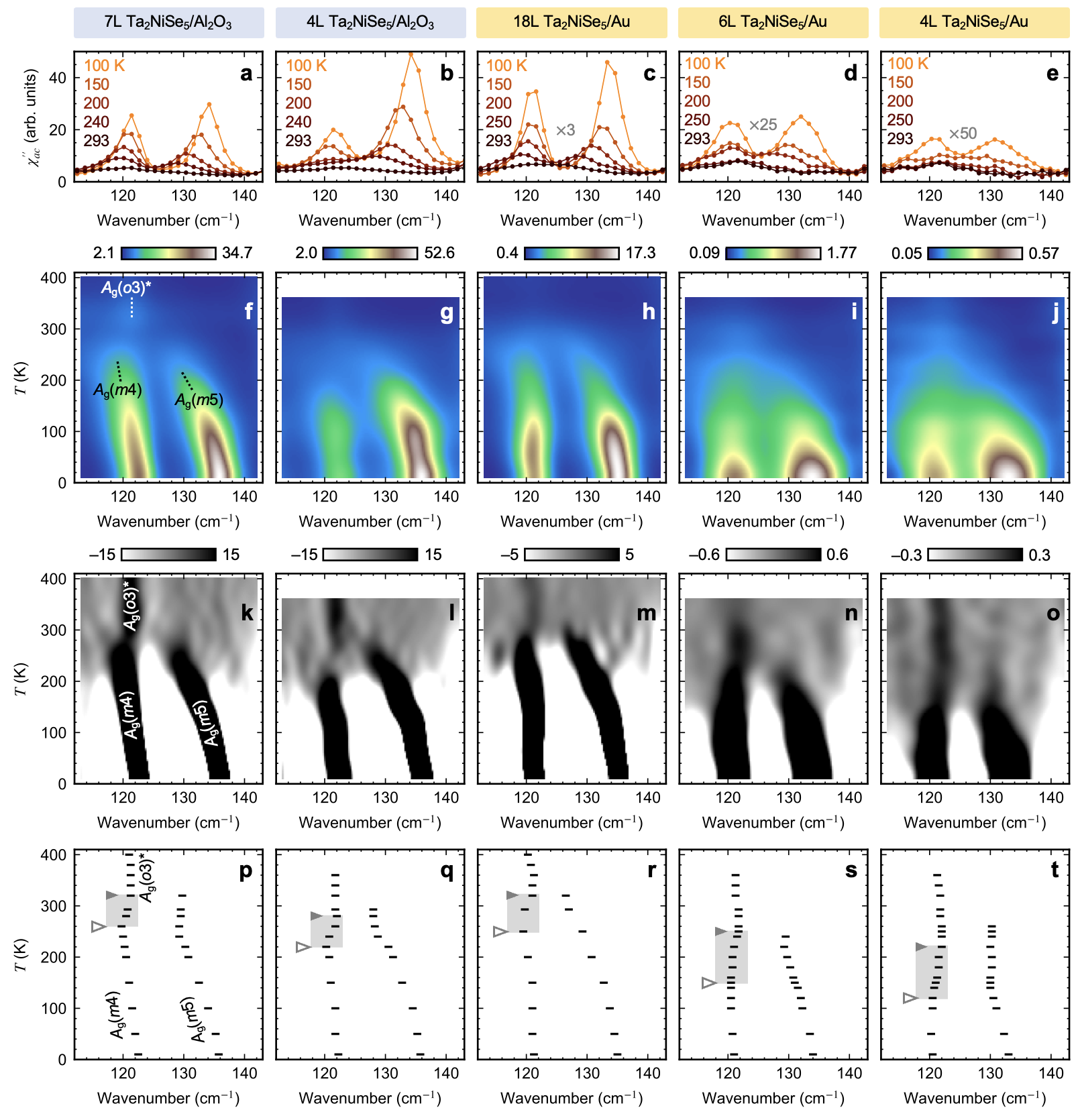}
\caption{\textbf{Detection of structural transition.} \textbf{a}--\textbf{e} $\chi''_{ac}(\omega)$ spectra at selected temperatures for Ta$_2$NiSe$_5$ flakes of various thicknesses on Al$_2$O$_3$ and Au. Note that the spectra in (\textbf{c})--(\textbf{e}) have been scaled by the indicated factors for comparison. \textbf{f}--\textbf{j} Interpolated 2D color maps of $\chi''_{ac}(\omega, T)$, showing phonon modes $A_{\textrm{g}}$($o$3)* (leaked from the $aa$ channel), $A_{\textrm{g}}$($m$4), and $A_{\textrm{g}}$($m$5). The scale bars have arbitrary units. \textbf{k}--\textbf{o} Plots of the second partial derivative $-\partial^2 \chi''_{ac} / \partial \omega^2$ corresponding to (\textbf{f})--(\textbf{j}). The scale bars have arbitrary units. \textbf{p}--\textbf{t} Peak positions of phonon modes $A_{\textrm{g}}$($o$3)*, $A_{\textrm{g}}$($m$4), and $A_{\textrm{g}}$($m$5). The width of the bars corresponds to the wavenumber resolution of the acquired data, 1.3 cm$^{-1}$. The solid triangles mark a kink in the temperature-dependent peak positions, which coincides with the onset of a monoclinic distortion, i.e., $T_{\textrm{c}}$. The open triangles mark a second kink at lower temperatures. From the temperature difference of the first and second kinks (vertical gray bars), we estimate the relative broadening of the structural transition in (\textbf{s}) and (\textbf{t}).}
\label{Fig3}
\end{figure}

Having confirmed the ability of polarized Raman spectroscopy to detect an orthorhombic-to-monoclinic distortion in Ta$_2$NiSe$_5$ flakes as thin as four layers, we proceed to investigate the structural transition in five samples: 7L and 4L flakes on Al$_2$O$_3$, and 18L, 6L, and 4L flakes on Au. We focus on the onset of phonon modes $A_{\textrm{g}}$($m$4) and $A_{\textrm{g}}$($m$5) in the $ac$ channel, as this is consistently the most prominent transition signature associated with the monoclinic distortion. We plot the temperature-dependent $\chi''_{ac}$ in the wavenumber range 112--143 cm$^{-1}$, both as individual $\chi''_{ac}$ vs.~wavenumber ($\omega$) spectra at selected temperatures ($T$) (Fig.~\ref{Fig3}a--e), as well as interpolated 2D color plots $\chi''_{ac}(\omega, T)$ (Fig.~\ref{Fig3}f--j). Above 320 K, $\chi''_{ac}$ for all five samples shows a wide background in this wavenumber range, which is part of the wide $B_{2\textrm{g}}$($o$2) mode of the orthorhombic phase (Fig.~\ref{Fig2}c), as well as a small narrow mode around 121--122 cm$^{-1}$, which is a leakage of mode $A_{\textrm{g}}$($o$3) from the $aa$ channel due to a small misalignment of polarizer and analyzer (indicated by asterisk). Below 320 K, $\chi''_{ac}$ for the 7L and 4L flakes on Al$_2$O$_3$ and the 18L flake on Au shows a clear and sudden onset of two new peaks, which are modes $A_{\textrm{g}}$($m$4) and $A_{\textrm{g}}$($m$5) of the monoclinic phase (Fig.~\ref{Fig3}f--h). These two modes also emerge in $\chi''_{ac}$ for the 6L and 4L flakes on Au; however, the onset appears to be lowered in temperature and more gradual in nature (Fig.~\ref{Fig3}i, j).  

We can better track the aforementioned phonon modes across the structural transition by examining the second partial derivatives $-\partial^2 \chi''_{ac} / \partial \omega^2$ of the interpolated 2D color plots (Fig.~\ref{Fig3}k--o). For the 7L flake on Al$_2$O$_3$ and the 18L flake on Au, the leakage mode $A_{\textrm{g}}$($o$3)* shows a kink and shift to lower wavenumbers below $\sim$300 K, just as it disappears and gives way to the monoclinic mode $A_{\textrm{g}}$($m$4) (Fig.~\ref{Fig3}k--m). Simultaneously, the monoclinic mode $A_{\textrm{g}}$($m$5) shows a sharp onset at higher wavenumbers. Similar behaviors are observed for the 4L flake on Al$_2$O$_3$, except that the kink due to the transition between $A_{\textrm{g}}$($o$3)* and $A_{\textrm{g}}$($m$4) occurs at a slightly lower temperature (Fig.~\ref{Fig3}l). For the 6L and 4L flakes on Au, these transition features are not only pushed down to even lower temperatures, but noticeably broadened in temperature range (Fig.~\ref{Fig3}n, o). The transition between $A_{\textrm{g}}$($o$3)* and $A_{\textrm{g}}$($m$4) is more continuous in appearance, and the onset of $A_{\textrm{g}}$($m$5) is preceded by a faint tail for the 4L flake on Au.  

To quantify the structural transition temperature, we extract the peak positions of the phonon modes in Fig.~\ref{Fig3}p--t. As previously explained, the transition between modes $A_{\textrm{g}}$($o$3)* and $A_{\textrm{g}}$($m$4) is marked by a small but abrupt shift to lower wavenumbers. We define this kink as the temperature below which the peak position begins to decrease (solid triangles). Using this criterion, we determine $T_{\textrm{c}}$ = 320$^{+10}_{-14}$ K for 7L Ta$_2$NiSe$_5$/Al$_2$O$_3$ (Fig.~\ref{Fig3}p), $T_{\textrm{c}}$ = 280$^{+7}_{-10}$ K for 4L Ta$_2$NiSe$_5$/Al$_2$O$_3$ (Fig.~\ref{Fig3}q), and $T_{\textrm{c}}$ = 320$^{+10}_{-14}$ K for 18L Ta$_2$NiSe$_5$/Au (Fig.~\ref{Fig3}r). The upper and lower limits of $T_{\textrm{c}}$ are determined from the temperature resolution of our measurements and are defined by a half-spacing to the neighboring data points. The $T_{\textrm{c}}$ value of 320$^{+10}_{-14}$ K for 18L Ta$_2$NiSe$_5$/Au agrees with the known $T_{\textrm{c}}$ value of 326 K for bulk Ta$_2$NiSe$_5$, which matches our expectation that 18 layers of a vdW compound is practically a bulk sample.

For the 6L and 4L flakes on Au, we can similarly identify 250 and 220 K as the temperatures below which the peak position of $A_{\textrm{g}}$($o$3)* shifts to lower wavenumbers (solid triangles in Fig.~\ref{Fig3}s, t). The kinks here, however, are broadened and less defined. To estimate the overall broadening, we note that in the previous three flakes, mode $A_{\textrm{g}}$($m$4) in the monoclinic phase exhibits a second kink at lower temperatures, below which its peak position shifts to higher wavenumbers (open triangles in Fig.~\ref{Fig3}p--r). This second kink occurs consistently around 60 K below the first kink for those three flakes (vertical gray bars). In the case of the 6L and 4L flakes on Au, this second kink is again broadened, and we roughly determine its temperature to be around 150 and 120 K, respectively (open triangles in Fig.~\ref{Fig3}s, t). This second kink occurs 100 K below the first kink for these two flakes (vertical gray bars). Given that the temperature difference between the two kinks is 60 K for the previous three flakes with sharp transitions, we deduce that the structural transitions in the 6L and 4L flakes on Au are broadened by $\pm$40 K around $T_{\textrm{c}}$. We crudely estimate ``average'' transition temperatures of $T_{\textrm{c}}^{\textrm{avg}}$ $\approx$ 250$\pm$40 K for 6L Ta$_2$NiSe$_5$/Au and $T_{\textrm{c}}^{\textrm{avg}}$ $\approx$ 220$\pm$40 K for 4L Ta$_2$NiSe$_5$/Au. \textcolor{blue}{(See Supplementary Note 3 with Supplementary Figure 4 for additional analysis of the structural transition.)}   

\subsection*{Substrate and layer dependence of electronic transition}

\begin{figure}
\includegraphics[width=\textwidth]{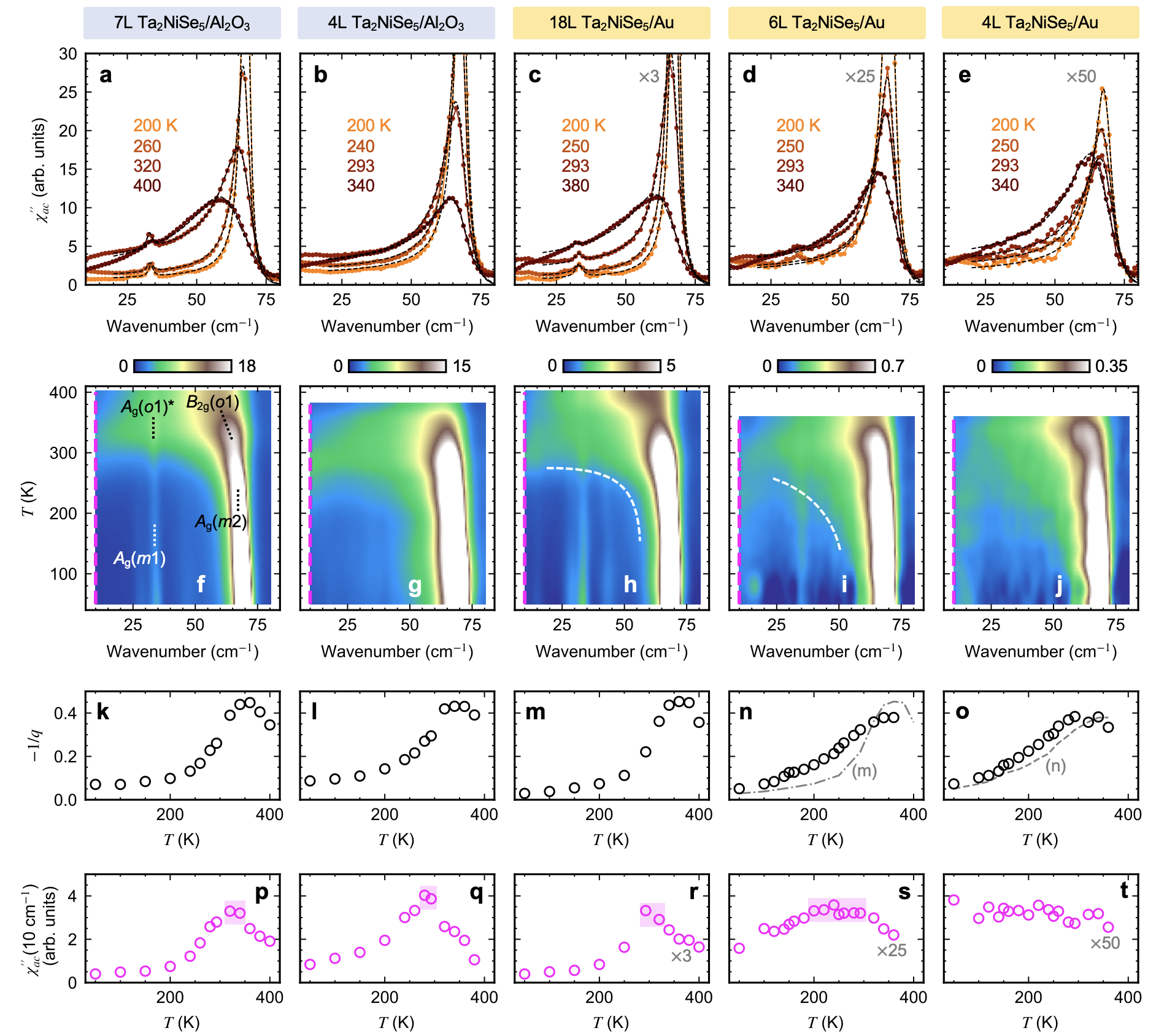}
\caption{\textbf{Detection of electronic transition.} \textbf{a}--\textbf{e} $\chi''_{ac}(\omega)$ spectra at selected temperatures for Ta$_2$NiSe$_5$ flakes on Al$_2$O$_3$ and Au. The spectra in (\textbf{c})--(\textbf{e}) have been scaled by the indicated factors. \textbf{f}--\textbf{j} Interpolated 2D color maps of $\chi''_{ac}(\omega, T)$, showing phonon modes $A_{\textrm{g}}$($o$1)* (leaked from the $aa$ channel), $B_{\textrm{2g}}$($o$1), $A_{\textrm{g}}$($m$1), and $A_{\textrm{g}}$($m$2). The dashed curves in (\textbf{h}) and (\textbf{i}) are guides to the eye. The scale bars have arbitrary units. (\textbf{k}--\textbf{o}) Fano asymmetry parameter $-1/q$ extracted from fits of $\chi''_{ac}(\omega)$ to (Eq.~\ref{Eq:Fano}) [dashed black lines in (\textbf{a})--(\textbf{e})]. The error bars from the fits are smaller than the circle symbols. In (\textbf{n}) and (\textbf{o}), the data from (\textbf{m}) and (\textbf{n}) are plotted for comparison, respectively (dashed/dotted lines). \textbf{p}--\textbf{t} Temperature-dependent low-energy spectra, i.e., $\chi''_{ac}(\textrm{10 cm}^{-1}, T)$ [taken along the dashed magenta lines in (\textbf{f})--(\textbf{j})]. The horizontal magenta bars include data points within 15\% of the peak value, from which we extract $T_{\textrm{c}}$.}
\label{Fig4}
\end{figure}

We turn our focus to Raman features that probe the electronic degrees of freedom in the Ta$_2$NiSe$_5$ thin flakes. Figure \ref{Fig4}a--e shows individual $\chi''_{ac}(\omega)$ spectra at selected temperatures in the wavenumber range 10--80 cm$^{-1}$. At temperatures of 340 K or higher, the dominant phonon mode $B_{2\textrm{g}}$($o$1) is broad and highly asymmetric with a long tail that extends down towards low wavenumbers. For the 7L and 4L flakes on Al$_2$O$_3$ and the 18L flake on Au, this mode (relabeled as $A_{\textrm{g}}$($m$2) in the monoclinic phase) quickly reverts to a sharper and more symmetric line shape as the temperature drops to 240--260 K (Fig.~\ref{Fig4}a--c). For the 6L and 4L flakes on Au, this transition to a sharper and more symmetric line shape appears to occur more slowly as temperature decreases (Fig.~\ref{Fig4}d, e). Figure \ref{Fig4}f--j visualizes these trends via interpolated 2D color plots of $\chi''_{ac}(\omega, T)$. In the displayed color scale, the green regions roughly span the long tail of the asymmetric mode $B_{2\textrm{g}}$($o$1). For the 7L and 4L flakes on Al$_2$O$_3$ and the 18L flake on Au, the green regions quickly collapse below the structural transition temperature, whereas for the 6L and 4L flakes on Au, the green regions shrink more gradually with lowering temperature (compare dashed guides in Fig.~\ref{Fig4}h, i).        

The asymmetric line shape of phonon mode $B_{2\textrm{g}}$($o$1), which is seen in $\chi''_{ac}$ above $T_{\textrm{c}}$, implies that this phonon is strongly coupled to a continuum, which in this case is an electronic continuum. Since $\chi''_{ac}$ of the bare Au and Al$_2$O$_3$ substrates are featureless, we attribute the origin of this electronic continuum to the Ta$_2$NiSe$_5$ flakes. In the orthorhombic phase of Ta$_2$NiSe$_5$, electronic Raman excitations with $B_{\textrm{2g}}$ symmetry can be ascribed to $Q$ $\approx$ 0 transitions between the Ni 3$d$ valence band and Ta 5$d$ conduction band, i.e., electron-hole excitations \cite{Volkov_2021}. To characterize the coupling strength between phonon mode $B_{2\textrm{g}}$($o$1) and this electronic continuum, we fit $\chi''_{ac}$ in the range 20--80 cm$^{-1}$ to a Fano function \cite{Fano_1961}, which describes the interference between a discrete resonance and a background: 
\begin{equation}
\chi''(\omega) = A\frac{(q + \epsilon)^2}{1+\epsilon^2},~~\epsilon = \frac{\omega - \omega_0}{\gamma}.
\label{Eq:Fano}
\end{equation}
Here, $\omega_0$ and $\gamma$ are the peak position and damping of the discrete resonance and $A$ controls the overall amplitude. (For the 7L flake on Al$_2$O$_3$ and the 18L flake on Au, we simultaneously fit the tiny peak between 30 and 35 cm$^{-1}$ from phonon mode $A_{\textrm{g}}$($o$1)*/$A_{\textrm{g}}$($m$1) to a Lorentzian.) We extract the Fano asymmetry parameter $–1/q$, which reflects the strength of electron-phonon coupling, as a function of temperature in Fig.~\ref{Fig4}k--o (sample fits are shown in Fig.~\ref{Fig4}a--e). For the 7L and 4L flakes on Al$_2$O$_3$ and the 18L flake on Au, $–1/q$ shows a maximum in the vicinity of $T_{\textrm{c}}$, followed by a rapid decrease upon lowering temperature (Fig.~\ref{Fig4}k--m). This behavior mirrors previous observations in bulk Ta$_2$NiSe$_5$ \cite{Kim_NatComm_2021} and coincides with the opening of a charge gap, which suppresses the low-energy electronic excitations that can couple to the phonon mode. For the 6L and 4L flakes on Au, the evolution of $–1/q$ with temperature appears to be broadened again, with a wider maximum at higher temperatures and a slower decrease at lower temperatures (Fig.~\ref{Fig4}n, o). We infer that the charge gap opens more gradually with decreasing temperature.

Another Raman feature indicative of an electronic transition is the temperature dependence of $\chi''_{ac}$ at low wavenumbers. In bulk Ta$_2$NiSe$_5$, the appearance of a low-energy spectral weight around $T_{\textrm{c}}$, i.e., a quasi-elastic peak, was reported and interpreted as either a direct signature of critical excitonic fluctuations \cite{Kim_NatComm_2021}, or an indirect signature of these excitonic fluctuations coupled to an acoustic phonon \cite{Volkov_2021}. Irrespective of the detailed microscopic explanation, the quasi-elastic peak provides us with a means to quantify the electronic transition temperature. Figure \ref{Fig4}p--t plots $\chi''_{ac}$(10 cm$^{-1}$) vs.~$T$ for our five flakes. For the 7L and 4L flakes on Al$_2$O$_3$ and the 18L flake on Au, $\chi''_{ac}$(10 cm$^{-1}$) vs.~$T$ shows a sharp peak at $T_{\textrm{c}}$, similar to the previous observations in bulk Ta$_2$NiSe$_5$. For the 6L flake on Au, the peak is significantly broadened, and for the 4L flake on Au, the peak is undetectable. By defining the electronic transition temperature and its width by the data points that lie within 15\% of the maximum value of $\chi''_{ac}$(10 cm$^{-1}$), we determine $T_{\textrm{c}}$ = 330$\pm$10 K for 7L Ta$_2$NiSe$_5$/Al$_2$O$_3$, $T_{\textrm{c}}$ = 287$\pm$7 K for 4L Ta$_2$NiSe$_5$/Al$_2$O$_3$, $T_{\textrm{c}}$ = 306$\pm$14 K for 18L Ta$_2$NiSe$_5$/Au, and $T_{\textrm{c}}^{\textrm{avg}}$ $\approx$ 246$\pm$46 K for 6L Ta$_2$NiSe$_5$/Au (horizontal magenta bars in Fig.~\ref{Fig4}p--s). These values agree with those derived from the structural transition (Fig.~\ref{Fig3}p--s). For the 4L Ta$_2$NiSe$_5$/Au flake, although we cannot detect a quasi-elastic peak, we emphasize that the existence of an electronic transition with charge-gap opening is confirmed by the temperature dependence of $-1/q$. By comparing $-1/q$ of the 6L and 4L flakes on Au on the same plot (see Fig.~\ref{Fig3}o), we observe that $-1/q$ of the 4L flake has similar temperature broadening as $-1/q$ of the 6L flake (dashed line), but is shifted $\sim$30 K lower in temperature. We therefore estimate the electronic transition of 4L Ta$_2$NiSe$_5$/Au to be $T_{\textrm{c}}^{\textrm{avg}}$ $\approx$ 220$\pm$46 K, using the same temperature width as that of 6L Ta$_2$NiSe$_5$/Au.   

\section*{Discussion}

\begin{figure}
\includegraphics[width=0.55\textwidth]{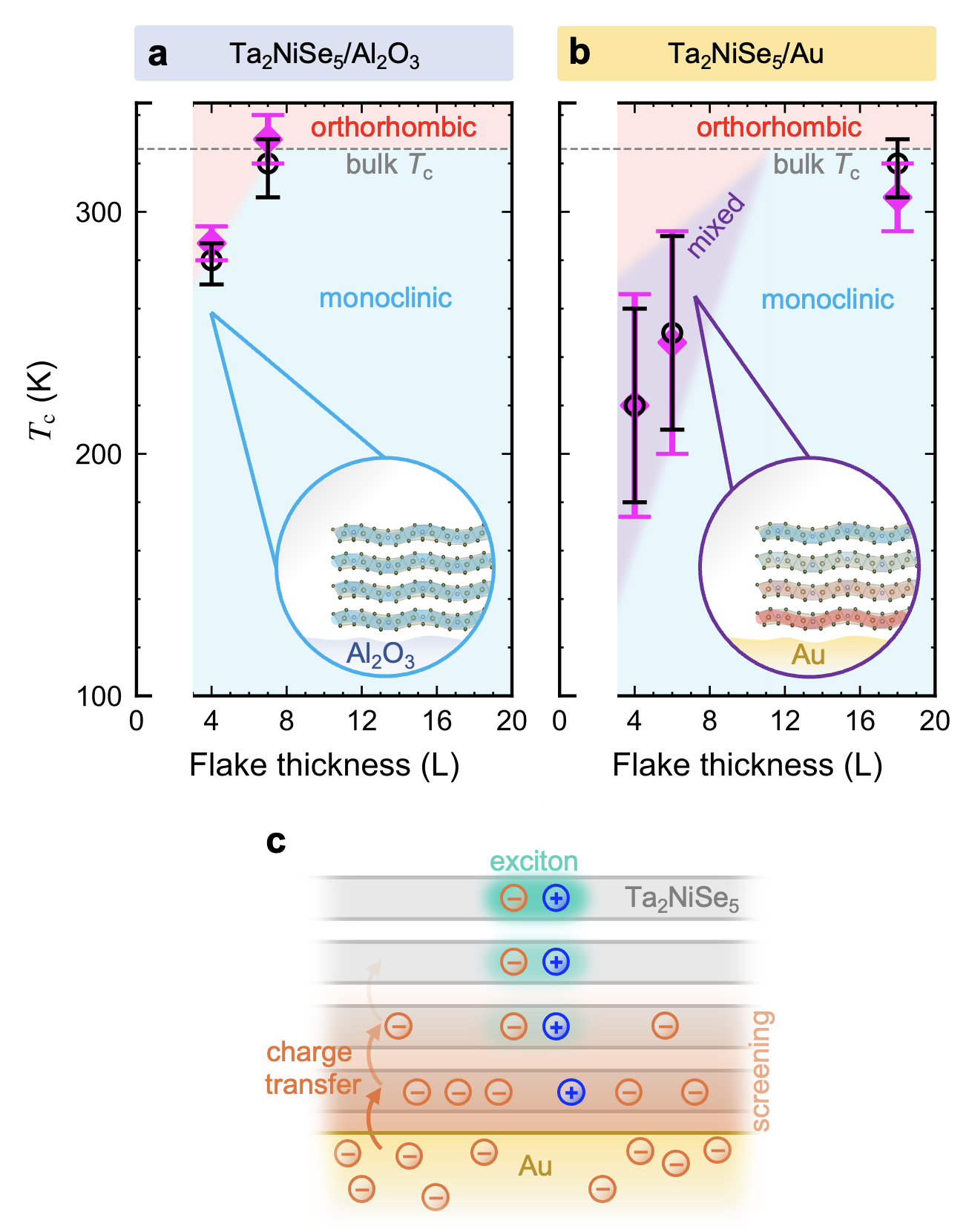}
\caption{\textbf{Phase diagram of Ta$_2$NiSe$_5$ thin flakes.} \textbf{a} Flakes on Al$_2$O$_3$ and \textbf{b} flakes on Au. The open circle symbols represent $T_{\textrm{c}}$ values extracted from the structural transition, with error bars determined from the analysis in Fig.~\ref{Fig3}p--t. The closed diamond symbols represent $T_{\textrm{c}}$ values extracted from the electronic transition, with error bars determined from the analysis in Fig.~\ref{Fig4}p--t. The dashed gray line represents $T_{\textrm{c}}$  = 326 K for bulk Ta$_2$NiSe$_5$. The inset cartoon in (\textbf{a}) shows a 4L Ta$_2$NiSe$_5$/Al$_2$O$_3$ flake in its monoclinic phase below $T_{\textrm{c}}$. The inset cartoon in (\textbf{b}) presents a microscopic picture for the broadened transition in 4L Ta$_2$NiSe$_5$/Au, where the layers closest to the Au substrate have a significantly lower $T_{\textrm{c}}$ due to an interface effect, whereas the surface layers have higher $T_{\textrm{c}}$. \textbf{c} Schematic cartoon illustrating the effects of charge transfer and screening on exciton binding close to the Ta$_2$NiSe$_5$/Au interface.}
\label{Fig5}
\end{figure}

Figure \ref{Fig5} summarizes our results for the structural and electronic transition temperatures of Ta$_2$NiSe$_5$ thin flakes of varying thicknesses on Al$_2$O$_3$ and Au. For the thin flakes on Al$_2$O$_3$, 7L Ta$_2$NiSe$_5$ is nearly bulk-like with a $T_{\textrm{c}}$ in agreement with 326 K, whereas 4L Ta$_2$NiSe$_5$ shows a $\sim$40 K decrease in $T_{\textrm{c}}$ (Fig.~\ref{Fig5}a). For the thin flakes on Au, 18L Ta$_2$NiSe$_5$ is nearly bulk-like with a $T_{\textrm{c}}$ in agreement with 326 K, whereas 6L and 4L Ta$_2$NiSe$_5$ show structural and electronic transitions that are significantly lowered and broadened with respect to temperature (Fig.~\ref{Fig5}b). From our crude estimates of $T_{\textrm{c}}^{\textrm{avg}}$, it appears that the broadened transition in the 4L flake begins around 260 K and is completed around 180 K, which means that $T_\textrm{c}$ is maximally suppressed by $\sim$150 K from the bulk value of 326 K. We conclude that substrate choice is an effective means of tuning the properties of Ta$_2$NiSe$_5$: A 4L flake on Al$_2$O$_3$ is moderately perturbed from bulk-like properties, whereas a 4L flake on Au shows sizable reduction and broadening of its transition temperature. 

We explain the microscopic origin for the lowered and broadened phase transitions in the 6L and 4L flakes on Au. Our results reveal an interface effect between Ta$_2$NiSe$_5$ and Au, i.e., the Ta$_2$NiSe$_5$ layers closest to the Au substrate are strongly perturbed with a local $T_{\textrm{c}}$ that is suppressed by roughly 150 K, whereas the Ta$_2$NiSe$_5$ layers farther away from Au have a local $T_{\textrm{c}}$ closer to room temperature. For a sample as thick as 18L, the fraction of interfacial layers is too small to contribute to the Raman signal, and we measure the majority bulk-like layers. For the 6L and 4L samples, Raman scattering from the interfacial Ta$_2$NiSe$_5$ layers becomes detectable in the overall signal. The region in the phase diagram spanned by the large error bars of $T_{\textrm{c}}^{\textrm{avg}}$ lies above the reduced $T_{\textrm{c}}$ of the interfacial layers, but below the nearly bulk-like $T_{\textrm{c}}$ of the surface layers. This region represents a mixed phase of Ta$_2$NiSe$_5$, wherein the layers closest to the surface have undergone a monoclinic distortion with charge-gap opening, while the layers closest to Au are still orthorhombic and marginally semimetallic (inset of Fig.~\ref{Fig5}b). Consequently, the experimental Raman susceptibility, which averages over all layers, shows a gradual onset of the monoclinic phonon modes $A_{\textrm{g}}$($m$4) and $A_{\textrm{g}}$($m$5) in the $ac$ channel (Fig.~\ref{Fig3}s, t), as well as a gradual decrease in the asymmetry of the $B_{2\textrm{g}}$($o$1) line shape, i.e., the Fano parameter $-1/q$, which reflects a gradual gapping out of low-energy electron-hole excitations (Fig.~\ref{Fig4}n, o). The fact that the transition broadening persists down to a 4L sample on Au implies that the interface effect from Au must be confined to one or two layers, i.e., not as large as four layers. The fact that the transition broadening is still observable in 6L Ta$_2$NiSe$_5$/Au is a testament to the sensitivity of Raman spectroscopy in detecting changes to the structural and electronic properties of interfacial layers that are buried $\sim$3 nm below the surface. \textcolor{blue}{(See Supplementary Note 4 with Supplementary Figure 5 for toy-model simulations of 4L Ta$_2$NiSe$_5$/Au with layer-dependent $T_\textrm{c}$ values.)}

The contrasting result that 4L Ta$_2$NiSe$_5$/Al$_2$O$_3$ shows a sharp structural and electronic transition just below room temperature implies the absence of any notable interface effect from Al$_2$O$_3$; otherwise, a mixed or broadened signal would have likewise been detected. Its reduced $T_\textrm{c}$ around 280--287 K is somewhat reminiscent of the reduced $T_\textrm{c}$ = 309 K previously reported for a 8L Ta$_2$NiSe$_5$ flake on insulating SiO$_2$ \cite{Kim_ACSNano_2016}. The sharp transition in 4L Ta$_2$NiSe$_5$/Al$_2$O$_3$ further serves as a control to exclude a degradation effect, in which the Ta$_2$NiSe$_5$ oxidizes and results in a lower $T_{\textrm{c}}$, as a possible origin for the broadening in 4L Ta$_2$NiSe$_5$/Au. Given the existence of an interface effect in Ta$_2$NiSe$_5$ on conducting Au but not on insulating Al$_2$O$_3$, we conclude that charge transfer and electrostatic screening from a conducting substrate are the likely culprit of the interface effect. While dielectric screening and defect-mediated charge transfer are still possible in Al$_2$O$_3$, they should be much weaker compared to the screening and charge transfer of free electrons in a highly metallic material like Au. 

Figure \ref{Fig5}c presents a schematic cartoon of how screening and charge transfer create a $T_\textrm{c}$ gradient across 4L Ta$_2$NiSe$_5$/Au within an EI picture. For the Ta$_2$NiSe$_5$ layers closest to the Au substrate, screening (orange shading in Fig.~\ref{Fig5}c) from mirror charges in Au weakens the Coulomb interaction (green shading) between electrons and holes, which destabilizes the binding of excitons and results in a large reduction of the local $T_\textrm{c}$. For the Ta$_2$NiSe$_5$ layers farther away from the Au substrate, the screening decreases, leading to the recovery of exciton binding and a local $T_\textrm{c}$ that is closer to the bulk value. Similarly, charge transfer is known to be attenuated across layers of a vdW material, and doping from an interface or surface is often limited to one or two layers \cite{Kim_NatComm_2016, Zheng_PRB_2016}. Here, the extra electrons that are donated from Au to the interfacial Ta$_2$NiSe$_5$ layers suppress the local $T_\textrm{c}$ either by acting as point-like impurities that break electron-hole pairs \cite{Zittartz_1967}, or by contributing to an overall increase of the carrier density, which again enhances the screening of Coulomb interactions \cite{Jerome_1967}. We note that both screening and charge transfer can lead to a broadened suppression of $T_\textrm{c}$, and our measurements do not allow us to disentangle the two effects.      

Furthermore, we observe in Fig.~\ref{Fig4}r, t that the conducting Au substrate has a sizeable effect on the quasi-elastic peak, not only broadening this peak, but suppressing its magnitude beyond detection in the 4L flake. This suppression affirms the nature of the quasi-elastic peak as being related to critical excitonic fluctuations, which would be destroyed by charge transfer and screening from Au. In contrast, we observe in Fig.~\ref{Fig4}m--o that although the conducting Au substrate broadens the temperature dependence of the Fano parameter for phonon mode $B_{2\textrm{g}}$($o$1), and shifts its maximum value to lower temperatures, it does not strongly reduce its overall magnitude. The maximum value of $-1/q$ is 0.38 for the 6L and 4L flakes on Au, which represents a 16\% reduction from the maximum value of 0.45 for the 18L flake on Au. This moderate reduction could be explained by a simple temperature broadening of the peak in $-1/q$. We therefore infer that the primary role of Au is not to weaken electron-lattice interactions, as represented by the Fano-like coupling of phonon mode $B_{2\textrm{g}}$($o$1) to an electronic continuum, but instead to weaken electron-hole interactions, as represented by the critical excitonic fluctuations at low energies. Based on $T_{\textrm{c}}^{\textrm{avg}}$ $\approx$ 220$\pm$40 K for 4L Ta$_2$NiSe$_5$/Au, we propose that the suppression of excitonic mechanisms lowers the 326 K $T_\textrm{c}$ in bulk Ta$_2$NiSe$_5$ by up to $\sim$150 K. Our estimate is consistent with other Raman estimates of the excitonic temperature scale in bulk samples, which take 137 K as a lower bound \cite{Kim_NatComm_2021, Volkov_2021}. The remaining $\sim$180 K or so in the 326 K $T_\textrm{c}$ of bulk Ta$_2$NiSe$_5$ could be ascribed to a combination of lattice contributions and electron-lattice coupling effects. Our results confirm the existence of excitonic effects that contribute to the phase transition in Ta$_2$NiSe$_5$, while acknowledging a sizable remnant $T_\textrm{c}$ that involves the lattice and its interactions with electrons and holes.

Our results utilizing conducting Au substrate as a means of charge-tuning Ta$_2$NiSe$_5$ is broadly consistent with previous reports of K-dosed Ta$_2$NiSe$_5$ \cite{Fukutani_PRL_2019, Chen_PRB_2020, Chen_PRR_2023, Lee_ACS_Nano_2024}. Angle-resolved photoemission spectroscopy measurements have shown a concurrent closing of the charge gap and appearance of a semimetallic state with increasing electron carriers, but without direct information on the lattice degrees of freedom. The comparison of our results with hole-doped Ta$_2$NiSe$_5$, however, is less clear. While bulk crystals of Ta$_2$NiSe$_5$ with Ni replaced by $p$-type Co show a large reduction in $T_\textrm{c}$ that exceeds 100 K \cite{Hirose_2023}, recent experiments on gate-tuned Ta$_2$NiSe$_5$ flakes have instead reported that $T_\textrm{c}$ is modestly enhanced by 5 K upon field-effect hole doping \cite{Wei_NatComm_2025}. Based on our present results, we cannot offer further insights here, but future measurements using conducting substrates that have a higher work function than Ta$_2$NiSe$_5$ ($\approx$5.4 eV \cite{Mor_2017}) may potentially allow us to access the hole-doped regime through an interface effect.    

\section*{Conclusions}

In summary, we have succeeded in tuning the phase transition of thin flakes of the EI candidate Ta$_2$NiSe$_5$ via a substrate effect, and detected changes to the structural and electronic $T_{\textrm{c}}$ originating from the interface using polarized Raman spectroscopy. The strong sensitivity of Ta$_2$NiSe$_5$ layers to conducting Au not only supports an excitonic component as one of the contributing factors to the phase transition, but may also enable patterned metallic electrodes, deposited either below or above individual Ta$_2$NiSe$_5$ layers, to define contrasting regions of orthorhombic and monoclinic Ta$_2$NiSe$_5$ and probe, e.g., the possible tunneling of excitons across both vertical and lateral junctions. The effect of Au on interfacial Ta$_2$NiSe$_5$ layers may also be an important consideration for nanoscale devices that harness Ta$_2$NiSe$_5$ for photodetection \cite{Li_2016, Qiao_2021, Zhang_2021, Liu_2024} and terahertz amplification \cite{Haque_2024}. More generally, our transfer protocol for thin flakes based on the cold welding of evaporated Au should enable high-quality, vdW materials to be prepared on a large variety of substrates, thereby enlarging the rich possibilities of interface engineering in vdW materials.    

\section*{Methods}

\subsection*{Sample fabrication}

Single crystals of Ta$_2$NiSe$_5$ were synthesized via chemical vapor transport, using I$_2$ as the transport agent \cite{Lu_NatComm_2017}. 

For the preparation of Ta$_2$NiSe$_5$ flakes, 20--30 nm of Au were thermally evaporated onto a wafer (SiO$_2$/Si or sapphire) and a Ta$_2$NiSe$_5$ crystal fixed on a piece of tape (Scotch or Nitto 224PR blue tape). The evaporation was carried out using a vacuum deposition system (MINIvap from MBRAUN) with base pressure $<$$5 \times 10^{-7}$ hPa, which was housed inside an inert Ar glove box ($<$0.1 ppm H$_2$O, $<$0.1 ppm O$_2$). Au pellets (3N5\%) were sublimated from a W boat at a deposition rate of 0.5--1 \AA/s. To facilitate the binding of Au onto the wafer (not Ta$_2$NiSe$_5$), a Cr adhesion layer of $\sim$2 nm was deposited onto the wafer prior to the evaporation of Au, using the same vacuum deposition system. The Au-covered Ta$_2$NiSe$_5$ on tape and the Au-covered wafer were then pressed and held together at room temperature, i.e., cold welded, before being peeled apart to create ultrathin flakes. All cold-welding and peeling processes were done inside the glove box. The cold welding was performed immediately after the Au deposition, whereas the peeling process could be performed up to weeks later, without any difference in the quality of flakes. \textcolor{blue}{Additional schematics and photos are provided in Supplementary Note 1 and Supplementary Figure 1.}

For the flakes on Al$_2$O$_3$, 50--100 nm of Al$_2$O$_3$ was first deposited onto the Ta$_2$NiSe$_5$ crystal via atomic layer deposition. The atomic layer deposition was performed in a chamber from Cambridge Nanotechnology Savannah (Veeco). Trimethylaluminum (TMA, Sigma-Aldrich Chemie GmbH) and H$_2$O were used as the precursors. The typical deposition temperature was 100--150~$^{\circ}$C, while the general pulse duration and purge duration were 10--200 ms and 5--30 s, respectively. Afterwards, the Al$_2$O$_3$-coated Ta$_2$NiSe$_5$ was transferred into our MINIvap chamber inside an Ar glove box, where Au was subsequently deposited as previously described.   

To confirm the thicknesses of the flakes, we used an atomic force microscope from Bruker in the PeakForce Tapping mode (see Supplementary Notes 5 and 6 with Supplementary Figures 6--8 and Supplementary Table 2 for further details).

\subsection*{Raman measurements and analysis}
	
Polarized Raman spectroscopy was performed using the 632.8 nm excitation line of a HeNe laser and a single-grating Jobin Yvon LabRAM HR800 spectrometer (HORIBA). For measurements below 300 K, the samples were loaded into a cryostat equipped with a sapphire viewport. To account for the attenuation from the viewport, a calibration factor was divided from the measured intensity $I$. The laser power was kept around 1 mW and we checked for the actual temperature by comparing the ratio of the Stokes and anti-Stokes signals. The spectra were acquired with a fixed wavenumber spacing of 1.3 cm$^{-1}$ and a variable temperature spacing ranging from 10 K (close to the transition temperature) to 50 K (far away from the transition temperature; e.g., below 100 K). The temperature spacings are shown in the data points plotted in Fig.~\ref{Fig3}p--t and Fig.~\ref{Fig4}k--t.

The intensity $I$ of the Stokes signal was converted into a susceptibility $\chi''$ via
\begin{equation}
\chi'' = \frac{I}{n(\omega, T) + 1},
\label{Eq:Bose}
\end{equation}
where $n(\omega, T) = 1 / [\exp(\hbar\omega/k_{\textrm{B}}T) – 1]$ is the Bose factor with the reduced Planck’s constant $\hbar$, frequency $\omega$, Boltzmann’s constant $k_{\textrm{B}}$, and temperature $T$. To remove a constant background due to the dark current of the charge-coupled device in the spectrometer, we subtracted the minimum intensity between 300 and 400 cm$^{-1}$, where the Raman signal is featureless, prior to converting $I$ into $\chi''$. 

To generate the 2D color plots shown in Fig.~\ref{Fig3}f--j and Fig.~\ref{Fig4}f--j, we interpolated our Raman susceptibility spectra $\chi''_{ac}(\omega, T)$ and resampled them on a 2D grid with 0.25 cm$^{-1}$ spacing in wavenumber and 5 K spacing in temperature. We then applied Gaussian smoothing with standard deviations of 1--2 cm$^{-1}$ in wavenumber and 10--15 K in temperature. In Fig.~\ref{Fig4}d, e, i, j, a spurious peak at 15 cm$^{-1}$ due to a leakage of the Rayleigh signal was removed from individual spectra (6L and 4L flakes on Au) and replaced via interpolation.

Figure~\ref{Fig3}k--o was generated by applying the partial second derivative $-\partial^2 \chi''_{ac} / \partial \omega^2$ to the 2D color plots in Fig.~\ref{Fig3}f--j. The peak positions in Fig.~\ref{Fig3}p--t were extracted from local maxima in the $-\partial^2 \chi''_{ac} / \partial \omega^2(\omega)$ spectra at selected measurement temperatures.  


\bibliography{References_Ta2NiSe5}

\subsection*{Acknowledgements}

We would like to thank G.~Blumberg, B.~Keimer, and L.~Wang for fruitful discussions and M.~Dueller, K.~Fink, A.~G{\"u}th, K.~K{\"u}ster, M.~Luo, T.~Reindl, A.~Schulz, J.~Smet, U.~Starke, and J.~Weis for technical support.

\subsection*{Funding}

This work was partly supported by the Deutsche Forschungsgemeinschaft (DFG, German Research Foundation) -- TRR 360 -- 492547816.

\subsection*{Author contributions}

M.I. synthesized the single crystals. Y.-S.Z. fabricated the thin flakes under the supervision of D.H. and H.T. Y.-S.Z. and C.X. performed the atomic layer deposition of Al$_2$O$_3$ substrates. Z.Y. and Y.-S.Z. performed the Raman measurements under the supervision of M.M. Y.-S.Z. and D.H. analyzed the data and wrote the manuscript with input from all coauthors.   

\subsection*{Competing interests}

The authors declare that they have no competing interests.

\subsection*{Data availability}

The data that support the findings of this study are available on Zenodo 

(DOI:~10.5281/zenodo.22107443).

\end{document}